\documentclass{sigplanconf}

\usepackage{amsmath}
\usepackage{graphicx}
\usepackage{hyperref}

\begin{document}

\setlength{\pdfpageheight}{\paperheight}
\setlength{\pdfpagewidth}{\paperwidth}

\conferenceinfo{CONF 'yy}{Month d--d, 20yy, City, ST, Country} 
\copyrightyear{20yy} 
\copyrightdata{978-1-nnnn-nnnn-n/yy/mm} 
\doi{nnnnnnn.nnnnnnn}




\titlebanner{banner above paper title}        
\preprintfooter{short description of paper}   

\title{Madeup: A Mobile Development Environment for Programming 3-D Models}
\subtitle{}

\authorinfo{Chris Johnson}
           {University of Wisconsin, Eau Claire}
           {johnch@uwec.edu}

\maketitle

\begin{abstract}
Constructionism is a learning theory that states that we learn more when we construct tangible objects. In the process of building and presenting our work, we make concrete the abstract mental models we've formed, see where they breakdown through the feedback we receive, and revise the models accordingly. Computer programming has long been taught under a constructionist approach using sensory-rich contexts like robots, media, and Logo-style environments. Now, with affordable 3-D printers in the hands of consumers, we have a new medium in which learners may realize their computational ideas. In this demonstration, we share a mobile development environment named Madeup, which empowers its users to navigate 3-D space using a Logo-like imperative and functional language. Every stop in space becomes a vertex in a 3-D model. The generated models may be exported or uploaded to a 3-D printing service.
\end{abstract}




\section{Description}
Madeup is an integrated development environment for programmatically generating 3-D models. Its primary purpose is to serve as a platform for teaching computation and algorithmic thinking to young learners of mathematics and computer science. The interface is comprised a text-based code editor and a 3-D canvas. Users program their models in a custom but traditional language that supports variables, operators, control structures, and modular abstraction. Similar to Logo's 2-D navigation commands, functions for moving and orienting in 3-D space are provided. When a Madeup program is executed, the path that is traced out by the navigation commands forms a 3-D model that may be exported or sent to a 3-D printer.

The Madeup environment is under active development by a group of computer science and mathematics educators. Given that its first formal evaluation will be conducted in a classroom equipped with mobile devices and that we want its general availability to be maximized, we are targeting mobile platforms.

\subsection{Learning}
At present, we are drawn to mobile devices primarily because they provide a portable learning environment. We'd like learners to have a chance to construct models while riding the bus or waiting in line. We will look at incorporating the other features of mobility as development progresses. For example, we have considered but not fully addressed the challenges of entering code without a hardware keyboard. Currently we eliminate the need for a hardware keyboard by reducing the use of non-alphabetic characters in the syntax of our language. For example, function calls do not require parentheses and parameters need only be delimited when they consist of more than one token.

Some non-alphabetic characters are precious---like carets for exponentiation and forward-slashes for division. Substituting other symbols that are easier to type would violate the sanctity of conventional mathematical notation. For this reason users are offered a palette of operators through the devices' automatic suggestion interface.

Madeup is designed to be used in a constructionist learning environment, where building and learning necessarily happen simultaneously. Constructionism is a theory of learning posited by Seymour Papert: ``The word constructionism is a mnemonic for two aspects of the theory of science education underlying this project. From constructivist theories of psychology we take a view of learning as a reconstruction rather than as a transmission of knowledge. Then we extend the idea of manipulative materials to the idea that learning is most effective when part of an activity the learner experiences as constructing is a meaningful product.''~\cite{papert87construction}.

Constructionism does not demand that the object being built is actually physically manifested, though Madeup certainly makes printing the created objects feasible. Instead, Papert argues that the most import quality of the learner's creation is that it is publicly perceivable~\cite{harel1991cn}. A long term goal of our project is to provide an online community for sharing code, objects, and lessons.

\subsection{Madeup Language}
The Madeup language is inspired by Logo. Though imperative, it provides several features of functional languages that support brevity of expression:
\begin{itemize}
\item Function definitions form closures. Variables defined in the enclosing scope at the time of definition do not need to be explicitly passed to the function. There is no concern that other code will alter the variables, as they are not dynamically scoped.
\item Types are determined dynamically. This obviates the need to declare variables. In a future iteration, we hope to implement static but implicit typing so that errors are detected earlier.
\item Every construct is an expression and has a value. A block, for example, evaluates to its last statement, as in Ruby and Scala. No explicit {\tt return} statement is needed. Conditional statements evaluate to the value of the block that is executed. Loops evaluate to the value of the repeated block's last iteration.
\end{itemize}

The Madeup grammar is defined using ANTLR~\cite{antlr}. The ANTLR-generated parse tree is traversed to build up an abstract syntax tree. The {\tt move} and {\tt moveto} commands emit vertex definitions. As vertices are emitted, geometry is assembled and sent to the mobile device's graphics processing unit, where it is shaded and placed on screen. Users may freely rotate their objects and turn on color or wireframe presentation styles to help them debug.

Constructing geometry through code is done in several different ways, which range from simple to complex. We expect users to begin by generating linear paths, or polylines. Madeup generates a polytube along the traversed path. Once familiar with the environment, users move on to walking through 2-D parametric spaces to generate spheres, toruses, cones, and other 2-manifolds. In both these modes, vertices are connected implicitly by the Madeup software to form the model. Advanced users may wish to form the geometry more directly, which we allow through a manual triangle forming mode.

\begin{figure}
\begin{center}
\includegraphics[width=0.5\linewidth]{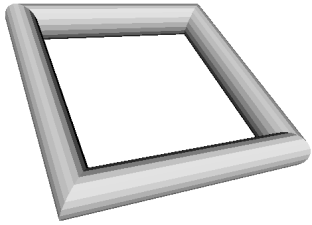}
\end{center}
\begin{verbatim}
repeat 4
  move 10
  yaw 90
end
\end{verbatim}
\caption{The canonical square frame, generated in polyline mode using a {\tt repeat} control structure.}
\label{fig:square}
\end{figure}

\begin{figure}
\begin{center}
\includegraphics[width=0.5\linewidth]{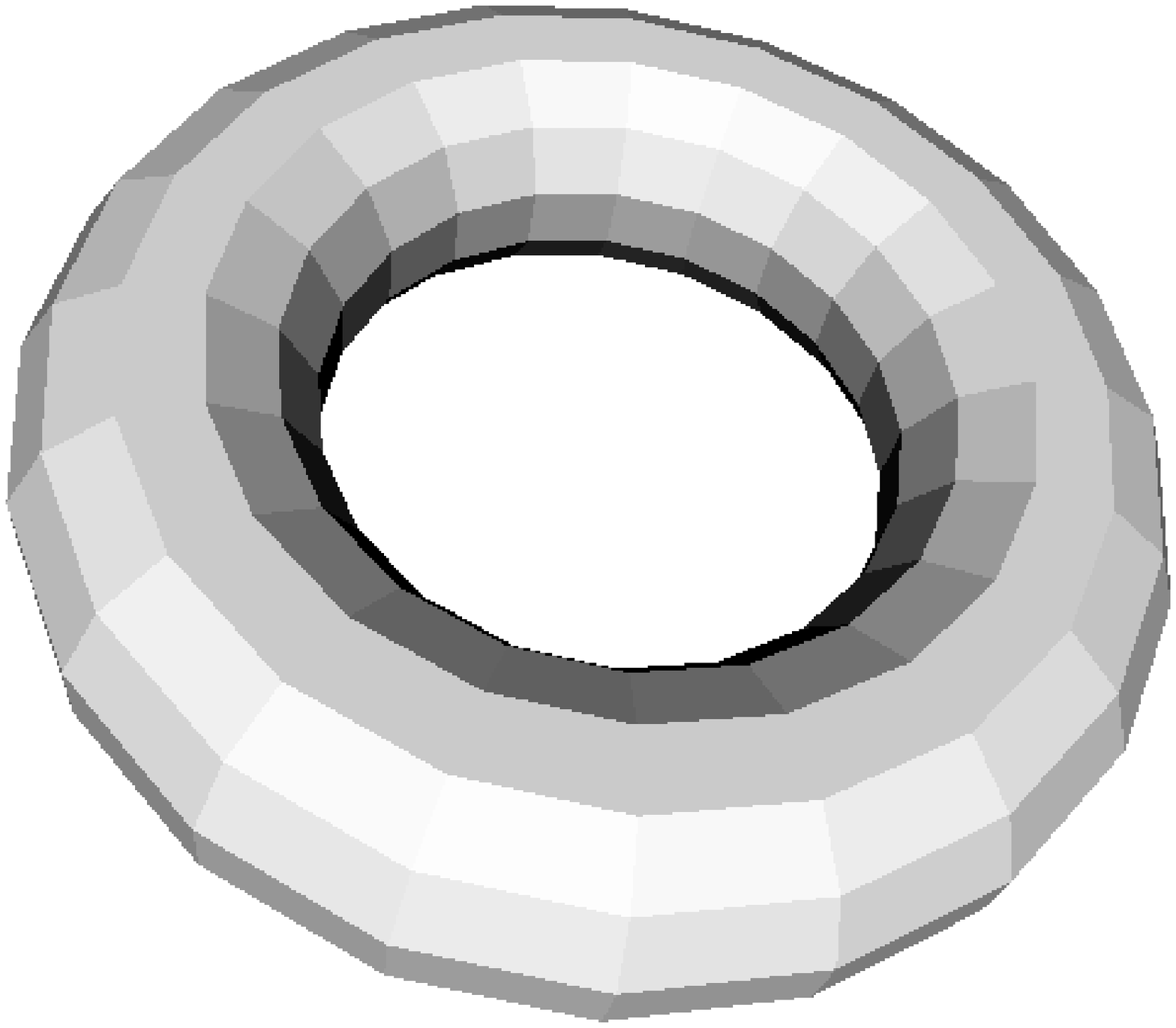}
\end{center}
\begin{verbatim}
nstops = 20
for i to nstops
  proportion = i / nstops * 2 * pi
  moveto (3 * sin proportion) 0 (3 * cos proportion)
end
\end{verbatim}
\caption{A discretized torus, generated in polyline mode by stopping 20 times along a circle of radius 3.}
\label{fig:torus}
\end{figure}

\begin{figure}
\begin{center}
\includegraphics[width=0.5\linewidth]{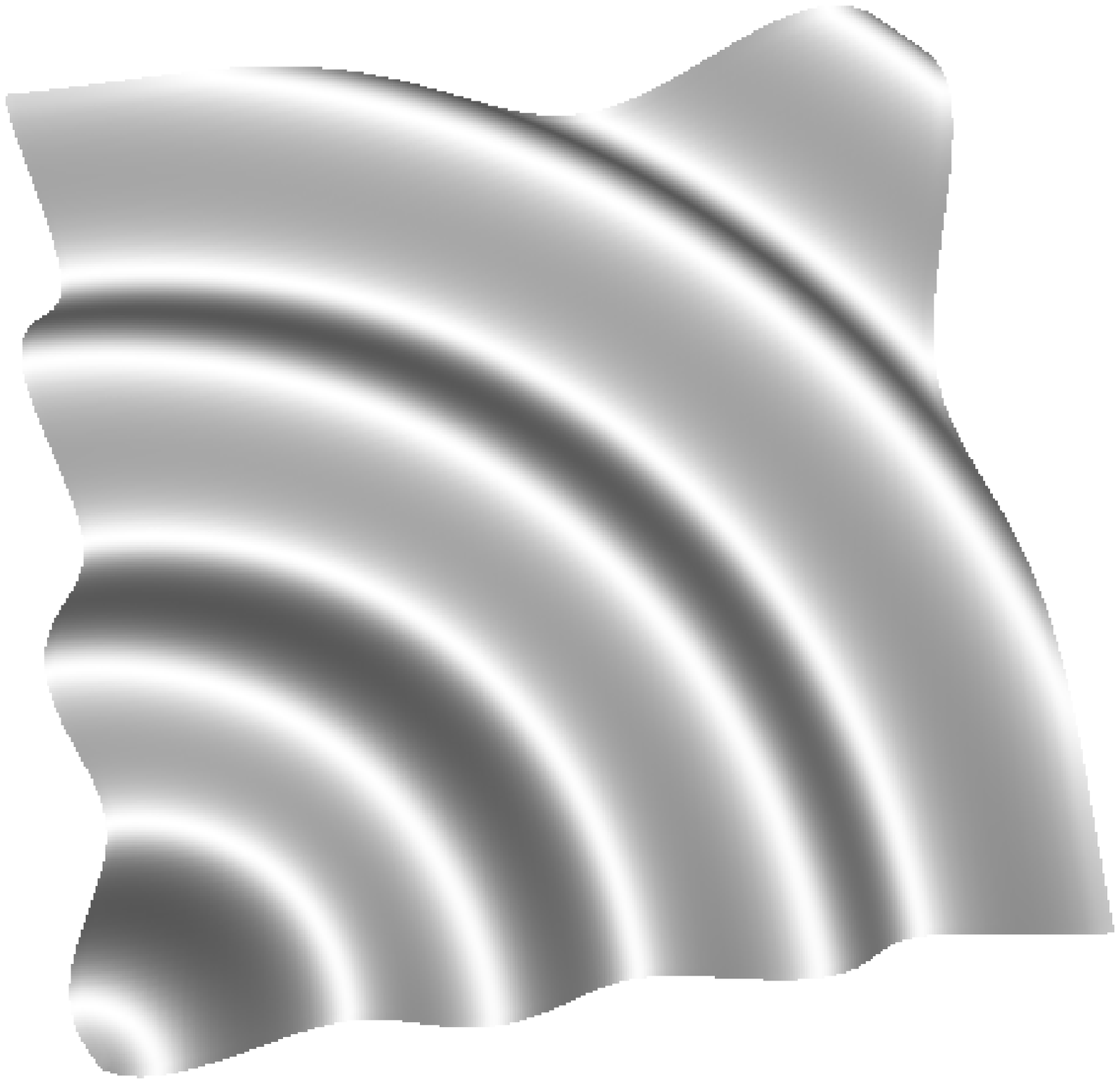}
\end{center}
\begin{verbatim}
length a b = (a * a + b * b) ^ 0.5
amplitude = 2
for r in 0..100
  for c in 0..100
    moveto c r (amplitude * sin(length c r))
  end
end
\end{verbatim}
\caption{A sinusoidal surface wave, generated in parametric mode. To reduce typing, actual parameters need not be delimited if they are comprised of bare identifiers or literals. Otherwise, parentheses are used to surround the subexpression.}
\label{fig:wave}
\end{figure}

Figures~\ref{fig:square} through~\ref{fig:wave} show several example Madeup programs and their accompanying 3-D results.

\subsection{Teaching Platform}
In addition to the sandbox environment that Madeup provides, it supports integrated playback of prerecorded interactive lessons. A teacher or peer may record a coding session, capturing both spoken annotations and every edit made to the code. The captured edits are preprocessed into a compressed form, where for each frame we store only the changes that were applied to the previous frame. We do this to reduce the size of the lesson, whose download may count against the learner's monthly data plan.

The movie is a ``text movie'' and is never locked up in pixels. The lesson is played back directly in the Madeup development environment, with the code appearing in the code editor and the results appearing in the 3-D canvas. Because the lesson code is text, the student may alter the content, scroll to other portions of the code to revisit context, and copy and paste useful snippets for their own projects.

In our current prototype system, lessons are recorded by instructors on a non-mobile platform. We have a medium-term goal of enabling recording of lessons directly in the Madeup environment. This would allow teachers and students to annotate their own creations and share them with others.

\acks
This work was supported in part by the SIGCSE Foundation through a Special Projects Grant.

\bibliographystyle{plain}


\begin{thebibliography}{1}

\bibitem{harel1991cn}
Idit Harel and Seymour Papert.
\newblock {\em Situating Constructionism}.
\newblock Ablex Publishing Corporation, 1991.

\bibitem{papert87construction}
Seymour Papert.
\newblock Constructionism: A new opportunity for elementary science education.
\newblock \url{http://nsf.gov/awardsearch/showAward?AWD_ID=8751190}.
\newblock [Online; accessed 15-August-2013].

\bibitem{antlr}
Terence Parr.
\newblock {ANTLR}: Another tool for language recognition.
\newblock \url{http://antlr.org}.
\newblock [Online; accessed 15-August-2013].

\end{thebibliography}




\end{document}